\documentclass{llncs}
\usepackage{amsmath}    
\usepackage{amssymb}    
\usepackage{mathtools}  

\usepackage{pgfplots}  
\pgfplotsset{compat=1.17}

\usepackage{algorithm}
\usepackage{algpseudocode}
\usepackage{array} 
\usepackage{float} 
\usepackage{caption} 
\usepackage{booktabs} 

\usepackage{mathrsfs}

\usepackage{stackengine}
\usepackage{url}

\begin{document}
\mainmatter              
\title{Computational Attestations of Polynomial Integrity Towards Verifiable Machine-Learning}
\titlerunning{CAPY2vML}  
%
\author{Dustin Ray\inst{1} \and Caroline El Jazmi\inst{1}}
\authorrunning{Ray, El Jazmi} 
%
\tocauthor{First Author, Second Author}
\institute{University of Texas Austin, Austin TX 78701, USA,\\
\email{dustinray@utexas.edu \& eljazmi@utexas.edu}
}

\maketitle              

\begin{abstract}
Machine-learning systems continue to advance at a rapid pace, demonstrating remarkable utility in various fields and disciplines. As these systems continue to grow in size and complexity, a nascent industry is emerging which aims to bring machine-learning-as-a-service (MLaaS) to market. Outsourcing the operation and training of these systems to powerful hardware carries numerous advantages, but challenges arise when privacy and the correctness of work carried out must be ensured. Recent advancements in the field of zero-knowledge cryptography have led to a means of generating arguments of integrity for any computation, which in turn can be efficiently verified by any party, in any place, at any time. In this work we prove the correct training of a differentially-private (DP) linear regression over a dataset of 50,000 samples on a single machine in less than 6 minutes, verifying the entire computation in 0.17 seconds. To our knowledge, this result represents the fastest known instance in the literature of provable-DP over a dataset of this size. We believe this result constitutes a key stepping-stone towards end-to-end private MLaaS.
\end{abstract}
\section{Introduction}
\label{sec:introduction}

\noindent The advent of cloud computing and software-as-a-service systems illustrates an unfolding trend in which many organizations have outsourced large aspects of their computing infrastructure to specialized external service providers, who in turn provide streamlined and robust access to networked hardware and software, often at a much lower cost than self-hosting such infrastructure. With the outsourcing of sensitive computing tasks, there arises a need for security infrastructure that allows for a consumer of such services to be confident that their information is protected and handled according to their specific needs.

Machine-learning, particularly generative systems, have continued to advance at a rapid pace, demonstrating utility in a wide variety of manners. A Deloitte study \cite{deloitte:genai} found that at least 50\% of surveyed organizations planned to use some form of machine-learning system in 2023. The growing utility of these systems has largely coincided with their ever-increasing complexity and dependence on vast troves of data used in the learning process. Training a machine-learning system is a computationally and financially expensive undertaking, which is often conducted using specialized hardware such as graphics processing units (GPUs). Access to this hardware is offered through a variety of commercial services, and it is the interaction with these services that presents security challenges for anyone who wishes to leverage a machine-learning system using outsourced hardware on sensitive training data.

Our penultimate result in this work shows how an MLaaS operator can employ novel cryptographic techniques with minimal hard assumptions to provide \textit{statements of computational integrity} to a consumer, such that:

\begin{itemize}
    \item The consumer is convinced with high probability that the work was carried out correctly.
    \item The verification of such a computation requires a proportionally small amount of work.
\end{itemize}

We argue that there is limited utility of an argument system which requires the consumer of this service to perform the same amount of work as the MLaaS operator, particularly during an ML training process. This research obtains an asymptotically optimal solution and applies this mechanism to a relatively open problem in privacy-preserving machine-learning (PPML) literature, which is the proving of correct application of differential-privacy during a training process. We discuss our design in depth in further sections.

The organization of this paper is structured as follows: Section \ref{sec:zkcrypto} discusses the fundamental principles of zero-knowledge (ZK) cryptography and computational integrity arguments, and presents a detailed account of our desired protocol attributes, emphasizing the requirements such as completeness, soundness, and succinctness that guide the design of our argument system. This section further explores the minimal hard assumptions necessary for the robustness of the protocol and provides the theoretical backbone for the argument systems employed to verify the integrity of computations in machine learning processes. 

Section \ref{sec:DP} formally defines differential-privacy and illustrates the application of the mechanism to a simple linear regression, and outlines the various differential privacy methodologies, comparing their applications and efficacy in maintaining privacy during data processing. This comparison underscores the selection of the appropriate differential privacy variant suited for our cryptographic arguments.

Section \ref{sec:zkprotocols} examines the different implementation approaches towards achieving efficient computational integrity statements in practice. Section \ref{sec:observations} shifts focus to the empirical evaluation of the proposed system, detailing the experimental setup, methodologies, and the performance metrics used to gauge the efficacy of the argument system in real-world scenarios.

Section \ref{sec:comparison} contains a literature review and analysis of the closest known result to ours. We use this result to highlight the contributions and advancements our approach offers over existing methods. The conclusion in Section \ref{sec:conclusion} encapsulates the significance of the research, its implications for privacy-preserving machine learning, and potential future directions.

\section{Zero-Knowledge Cryptography and Arguments of Computational Integrity}
\label{sec:zkcrypto}

\noindent This work draws from an active field of cryptography surrounding the study and application of argument systems, specifically those of the probabilistically-checkable and non-interactive variety. The celebrated results of \cite{gold:roth:shaf:yehu} realize an argument system consisting of a series of interactions between a prover $\mathscr{P}$ and a verifier $\mathscr{V}$, in which $\mathscr{P}$ tries to convince $\mathscr{V}$ of the truth of some statement, or, in a more applied context, that the output of a computation was obtained by computing over some input. \cite{gold:roth:shaf:yehu} shows that while $\mathscr{V}$ could re-run the entire computation and compare outputs with $\mathscr{P}$, a logarithmic-size, non-deterministic sampling of an argument provided by $\mathscr{P}$, will, with extremely high probability, be sufficient to satisfy $\mathscr{V}$ that the output was obtained correctly, and conversely otherwise. We define this protocol between $\mathscr{P}$ and $\mathscr{V}$ below.

\subsection{Protocol Attribute Constraints}
\label{sec:protocol}

\noindent In probabilistic polynomial time, we strive for the following properties for a robust and efficient argument system under random oracle assumptions: \cite{ben:chie:spoo}

\textit{Completeness}: True statements can always be proven by a prover and will always be accepted by a verifier, except with negligible probability. Formally: For every instance-witness pair $(\mathscr{x, w}$) in a relation $\mathscr{R}$, Pr $\left [\mathscr{V}^p \left ( \mathscr{x}, \mathscr{P}^p (\mathscr{x}, \mathscr{w})\right ) = 1  \right ] = 1$ for probability $p$ taken over $\mathscr{P}$ and any randomness from $\mathscr{P}$ or $\mathscr{V}$.

\textit{Soundness}: A prover should not be able to deceive a verifier into accepting a false statement as true, except with negligible probability. Formally: For every instance $\mathscr{x}$ not in the language of $\mathscr{R}$ and every malicious prover $\widetilde{\mathscr{P}}$ submitting at most a polynomial number of queries to a random oracle, $Pr \left [\mathscr{V}^p \left ( \mathscr{x}, \\ \mathscr{\widetilde{\mathscr{P}}}^p\right )= 1 \right ]$ is negligible in the security parameter. \cite{ben:chie:spoo} shows how a quantifiable lower-bound of security can be derived from the soundness parameter, which confers a configurable level of $n-$bit security directly from the security of the choice of underlying hash function.

\textit{Succinctness}: We define the relationship between the work done by $\mathscr{P}$ and $\mathscr{V}$ in our construction. Strictly speaking, if $\Omega(n)$ is the amortized asymptotic upper bound of complexity for argument generation and verification respectively, then for our construction, we strictly require that $\Omega(n)_\mathscr{P} \leq quasi(n)$ and for $\Omega(n)_\mathscr{V} \leq polylog(n)$. In other words, we restrict the upper bound on the work done by the prover as quasi-linear, and the work performed by the verifier as poly-logarithmic. This definition of succinctness leads to a construction uniquely suited to the MLaaS setting. An ML operator can perform a computationally expensive algorithm, the result of which can be verified for correctness in logarithmic time (and size) with respect to the computation itself. This is a significant result with important ramifications: a hypothetical computation requiring 10,000,000 steps can be verified with \textit{only 23 queries}.

\textit{Minimal Hard Assumptions}: Wherever possible, we desire the protocol to rely on minimal cryptographic hard assumptions. The result from \cite{ben:chie:spoo} satisfies this requirement through the use of only a secure hash function as the underlying primitive, then paired with error-correcting codes, which ultimately lead to plausibly post-quantum-secure computational integrity statements. We do not simulate a quantum adversary in this work, but the security of our scheme follows naturally from the underlying construction \cite{ben:chie:spoo} in use.

\textit{Perfect Zero-Knowledge}:  We make a subtle divergence from the nomenclature of \cite{ben:ben:hor:ria} in our usage of the term "computational integrity statements", by observing that such statements do not require perfect zero-knowledge in our scheme. ZK in our result follows trivially from the application of \cite{ben:ben:hor:ria} with surprisingly little overhead, but it is not necessarily a requirement for our protocol because the MLaaS operator and consumer are assumed in this setting to be the only parties involved in the computation, both having access to the same sensitive data and resulting machine-learning model. However, since the argument itself is zero-knowledge following the application of \cite{ben:ben:hor:ria}, it indeed reveals nothing of the model or dataset used in the computation, and could be passed to any public, untrusted party for safe and efficient verification.

The previously defined parameters for our protocol are realized through the application of a form of cryptographic non-interactive argument scheme resembling a \textit{zero-knowledge scalable transparent argument of knowledge} (ZK-STARK) \cite{ben:ben:hor:ria} to a machine-learning algorithm known as a \textit{differentially-private linear regression} \cite{alabi:mcmil:sar:smi:vad}.

\subsection{Nomenclature}

\noindent It is common in ZK literature to refer to the computational integrity statement as a "proof", and as the party generating the statement as the "prover". A protocol is characterized with desirable asymptotic traits when requirements around perfect soundness are relaxed. Hence, throughout this work, we refer to such statements as \textit{arguments}, because they do not equivocally prove the truth of a statement, but rather they \textit{argue} the truth of the statement. The term "zero-knowledge proof" is understood to be equivalent to, and implying meaning of, the term "computational integrity (CI) statement", or "argument", and thus we use these latter two terms throughout the work.

We refer to the party generating the statements as the prover. The honest prover does in fact posses an entire, complete proof that the given computation was carried out correctly. To achieve optimal asymptotic complexity, the prover will transform the proof into an argument. The term "prover" thus accurately reflects the role of the party who produces computational integrity statements, and as such we use this term throughout this work. In section \ref{sec:observations}, we often use the phrase "proving the dataset". This is shorthand indicating that we create a CI statement for the training process over the entire dataset.

Lastly, we prefer the terminology in \cite{gold:roth:shaf:yehu} with regards to how we obtain a linear regression over some dataset. \cite{gold:roth:shaf:yehu} derives this terminology from the study of "Probably Approximately Correct" (PAC) learning, which is a framework for the mathematical analysis of machine learning. We use the term "learn" and "hypothesis" to indicate machine-learning training and the resulting model respectively.

\section{Differential Privacy}
\label{sec:DP}

\noindent The literature arising from the intersectional disciplines of theoretical cryptography and applied machine-learning is rich with technical achievements in secure multi-party computation (MPC) protocol design towards the goal of realizing machine-learning (ML) systems that can be operated securely and privately. There are abundant \cite{lee:kif}, \cite{abadi:deep}, \cite{goog:ai:dp} examples of secure privacy mechanisms and constructions that can be readily leveraged in a variety of applications. This work advances the field by exploring the robust and efficient application of cryptographically provable privacy mechanisms towards machine-learning training processes.

Differential privacy is a mechanism applied solely by an MLaaS operator, which, in a standard setting, then has no immediate means of attesting that the mechanism actually was involved in subsequent computations as agreed to by the consumer. The best the consumer can do in this situation is place trust in a possibly malicious MLaaS operator that their data was trained over with the desired privacy mechanism in place. This work develops a means for the MLaaS operator to act as a prover, and the consumer to act as a verifier in an Interactive-Oracle-Proof (IOP) model. The consumer performs a small amount of work to verify that the result of training could only have been obtained by correct execution of an agreed-upon set of steps.

We begin by introducing \textit{differential privacy}, which we show can be provably applied during a machine-learning training process. Differential privacy is presented formally: 
\begin{definition}($\epsilon$, $\delta$) - Differential Privacy
\label{def:dif_priv}
\cite{dwo:rot}, \cite{fath:noise}:
    A randomized algorithm $\mathscr{M}$ with domain $\mathscr{N}^{|\mathscr{X}|}$ is ($\epsilon$, $\delta$) - differentially private if for all S $\subseteq$ Range($\mathscr{M}$) and for all $x, y \in \mathscr{N}^{|\mathscr{X}|}$ such that $||x-y||$$_{1}$ $\leq 1$:

    \begin{equation}
    \label{eqn:DP}
    \text{Pr}[\mathscr{M}(x) \in S] \leq \text{exp}(\epsilon)\text{Pr}[y \in S] + \delta
    \end{equation}

\end{definition}
Where: 

\begin{itemize}
    \item $\mathscr{M}$: A randomized algorithm (query(db) + noise, or query(db+noise))
    \item \textit{S}: All possible outputs of $\mathscr{M}$ that could be guessed
    \item $x$: Entries in database ($\mathscr{N}$)
    \item $y$: Entries in parallel database ($\mathscr{N} \pm 1$)
    \item $\epsilon$: Maximum distance between a query on $\mathscr{N}(x)$ and the same query on $\mathscr{N}(y)$
    \item $\delta$: The probability of some given information being leaked
\end{itemize}
We emphasize that definition 1 holds for a single query and \textit{not} for multiple queries, and that it does not imply that an algorithm is differentially private, rather it is a measure of how much privacy is leaked to an observer given a single query on a database. The notion of a parallel database $\mathscr{N} \pm 1$ is meant to signify a database that differs by a single entry from $\mathscr{N}$. Definition 2 then shows that $\epsilon$ and $\delta$ are measures of by how much the probability distributions of the entries of $\mathscr{N}$ differ from $\mathscr{N} \mp 1$.

\subsection{Epsilon $(\epsilon)$}

\noindent \cite{fath:noise} shows how $\epsilon$ is a metric of privacy loss at a differential change in a database, such as when an entry is added or removed. $\epsilon$ is defined as the maximum distance between a query on database $\mathscr{N}$ versus the same query on database $\mathscr{N} \pm 1$. It is necessary to examine the effect that differing values of $\epsilon$ has on the privacy of a given database query. 

$\delta$ is a value typically defined to be an exceedingly small bias that represents the possibility that some information is leaked from a given query over a database. It is common in literature to choose $\delta$ to be $||\mathscr{N}||^{-1}$, or the inverse of the size of the given database to be processed. 

($\epsilon$, $\delta=0$) differential privacy indicates that an adversary cannot distinguish whether the output of an algorithm $\mathscr{M}$ was produced by processing $\mathscr{N}$ versus $\mathscr{N} \pm 1$. As the value of $\epsilon$ approaches zero, the privacy guarantees offered by such values becomes increasingly similar. Conversely, larger values of $\epsilon$ indicate that there exists an adversary that can distinguish with higher probability whether the output of $\mathscr{M}$ was obtained from either $\mathscr{N}$ or $\mathscr{N} \pm 1$. $\epsilon$-DP thus facilitates control over how much privacy can be "added" to a given database query. Smaller values of $\epsilon$ require that queries over $\mathscr{N}$ and $\mathscr{N} \pm 1$ produce similar outputs. In the following sections, we show precisely how this can be achieved in terms of various probability distribution mechanisms.

\subsection{Laplace Mechanism}

\noindent A numeric database query can be defined as: 
\begin{center}
    $f: \mathscr{N}^{|\mathscr{X}|} \rightarrow \mathscr{R}^{k}$
\end{center}

\noindent  Such a query maps a database to $k$ real numbers. We proceed to define $\mathscr{l}1$ sensitivity, which is a measure of how a mapping will respond to adjacent datasets different by only a single entry:

\begin{definition}[$\mathscr{l}1$ sensitivity:]
\label{def:l1_sensitivity}
    \cite{dwo:rot}
     The $\mathscr{l}1$-sensitivity of a function $ f: \mathscr{N}^{|\mathscr{X}|} \rightarrow \mathscr{R}^{k}$ is:  
    \begin{equation}
    \label{sensitivity}
        \Delta f = \stackanchor{max}{\stackanchor{$x,y \in \mathscr{N}^{|\mathscr{X}|}$}{$||x-y||_{1}=1$}} ||f(x)-f(y)||_{1}.
    \end{equation}
\end{definition}

\noindent  Definition 2 provides a means of measuring the maximum effect of a single entry $x$ in a database on the output of the function $f$, which in turn helps to quantify the amount of noise that should be added to the output of $f$ in order to hide the presence of $x$ from the output of $f$. We proceed to define the Laplace distribution, which can be leveraged as a source of noise to incorporate into $f$. 

\begin{definition}The Laplace Distribution: The Laplace distribution centered at zero with scale $b$ is the distribution with the probability density function:
    \begin{equation}
    \label{laplace}
        \text{Lap}(x | \mu ,b )\frac{1}{2b}\text{exp}\left ( - \frac{|x - \mu |}{b} \right )
    \end{equation}
\end{definition}

\noindent Definition 3 defines the Laplace distribution as two symmetric exponential distributions with an additional location parameter \cite{wiki:laplace}. The Laplace mechanism then will run algorithm $f$ and perturb each input with noise drawn from the Laplace distribution. The particular amount of noise to be added to an input is obtained by calculating the $\mathscr{l}1$ sensitivity of $\frac{f(\text{query})}{ \epsilon}$, with the added condition that $\delta = 0$, and thus the Laplace mechanism achieves ($\epsilon$, 0) differential privacy.\\

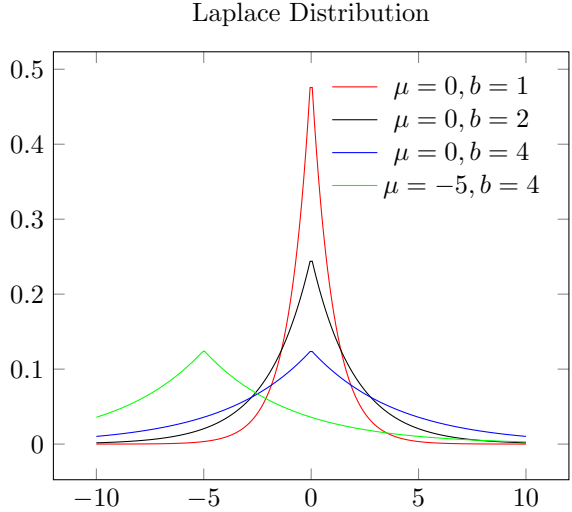
\begin{center} 
\begin{tikzpicture}
  \begin{axis}[
    samples=200,
    domain=-10:10,
    title=Laplace Distribution,
    legend style={fill=none, draw=none}, 
    legend pos=north east 
  ]
    \newcommand{\mean}{0} 
    \newcommand{\scale}{1} 
    
    \addplot[red] {1/(2*\scale) * exp(-abs(x - \mean)/\scale)};
    \addlegendentry{$\mu = 0, b=1$}
    
    \newcommand{\meanTwo}{0} 
    \newcommand{\scaleTwo}{2} 
    \addplot[black] {1/(2*\scaleTwo) * exp(-abs(x - \meanTwo)/\scaleTwo)};
    \addlegendentry{$\mu = 0, b=2$}

    \newcommand{\meanThree}{0} 
    \newcommand{\scaleThree}{4} 
    \addplot[blue] {1/(2*\scaleThree) * exp(-abs(x - \meanThree)/\scaleThree)};
    \addlegendentry{$\mu = 0, b=4$}

    \newcommand{\meanFour}{-5} 
    \newcommand{\scaleFour}{4} 
    \addplot[green] {1/(2*\scaleFour) * exp(-abs(x - \meanFour)/\scaleFour)};
    \addlegendentry{$\mu = -5, b=4$}

  \end{axis}
\end{tikzpicture}
\end{center}

\begin{definition}\label{def:l1_sensitivity}

The Laplace Mechanism \cite{dwo:rot}: Given any function $ f: \mathscr{N}^{|\mathscr{X}|} \rightarrow \mathscr{R}^{k} $, the Laplace mechanism is defined as:
    \begin{equation}
    \label{lap_mech}
        \mathscr{M}_L(x, f(\cdot), \epsilon) = f(x)+(Y_{1}...,Y_{k})
    \end{equation}
\end{definition}

\noindent Where all $Y_i$ are independently identically distributed random variables drawn from Lap $(\frac{\Delta f}{\epsilon})$ . Formally stated:
\\\\
\indent The Laplace Mechanism: Given any function $f: \mathscr{N}^{|\mathscr{X}|} \rightarrow \mathscr{R}^{k}$, the following definition of $F(x)$ satisfies $\epsilon$-differential privacy:

\begin{equation}
\label{alt_lap}
    F(x) = f(x) + \text{Lap}(\frac{S}{\epsilon})
\end{equation}

\noindent Where $S$ is the sensitivity of $f$, and $Lap(\frac{S}{\epsilon})$ denotes sampling from the Laplace distribution with center 0 and scale $S$. We use sensitivity in equations 4 and 5 to quantify the change in output from $f$ when the input database changes by exactly a single entry. We have thus presented a definition of differential privacy in terms of the Laplace mechanism. If we consider our problem space to be that of measuring the accuracy of computing $f$, the Laplace mechanism provides a general-purpose approach in which we can achieve differential privacy.%

\subsection{Differential Privacy Variations}

\noindent Differential-privacy is amenable to instantiation over varying hyperparameter choices. This allows for a choice of privacy budget which is suitable to the task at hand. In this context, our research focuses on the efficacy of a novel non-interactive argument system, specifically engineered to enhance both the integrity and verification mechanisms essential for the deployment of machine-learning algorithms under privacy constraints. Among the differential privacy variants evaluated, Differentially-Private Ordinary Least Squares (DP-OLS) is identified as the most appropriate privacy-preserving error estimation function for a linear regression. Other differential privacy methodologies, such as Pure Differential Privacy (Pure DP), Approximate Differential Privacy (Approximate DP) and Rényi Differential Privacy (RDP) were also considered.

{\textbf{Pure DP ($\varepsilon$-DP)}}, in which $\delta$ = 0, is known as the strongest version of differential privacy \cite{stei:ull}. Originating from a seminal paper by Dwork et al. \cite{dwo:mcs:nis:smi}, which introduced the initial definition of what we now recognize as Pure DP, this method adjusts the added noise based on the $\ell_1$ sensitivity of the query or queries being processed. This approach \cite{dwo:mcs:nis:smi} shows the incorporation of noise scaled to a privacy parameter, $\varepsilon$, ensuring that the probability of generating any particular output remains relatively unchanged even if the data of any single individual in the dataset is modified. Despite its strong privacy assurances, Pure DP typically necessitates the introduction of a higher level of noise \cite{das:mish}, which may disproportionately compromise the utility of simpler statistical models such as OLS. In the context of implementing DP-OLS, Pure DP could be applied by adding noise directly to the OLS coefficients. While the stringent privacy guarantees of Pure DP are well-suited for highly sensitive applications, the considerable noise added can often markedly diminish the utility of the regression model.

\textbf{Approximate DP ($(\mathbf{\varepsilon}, \mathbf{\delta})$-DP)}, while thoroughly investigated in scenarios where achieving Pure DP is complex, is less explored when the privacy loss parameter, $\delta > 0$ \cite{stei:ull}. Work by Bun, Ullman, and Vadhan \cite{bun:ull:vad} has identified strong lower bounds for this type of privacy, which are nearly optimal at $\delta \approx \frac{1}{n}$. This level represents the weakest privacy guarantee that still maintains practical relevance \cite{stei:ull}. In contrast, DP-OLS adheres to more rigid differential privacy mechanisms, closely resembling those used in Pure DP. The $(\epsilon, \delta)$ differential privacy model, characteristic of Approximate DP, introduces an additional parameter $\delta$, which denotes a small probability where the privacy guarantee might not be fully upheld \cite{wang:lei:fien}. For linear regression models, maintaining strong privacy without significantly affecting model accuracy is crucial. While Approximate DP may be preferable in scenarios where a minor relaxation of privacy is acceptable to gain computational efficiency in more complex models, DP-OLS stands out for its strong privacy guarantees.

{\textbf{Rényi DP ($(\alpha, \varepsilon)$-RDP)}}, represents a natural relaxation of the standard DP model, preserving many of its core properties while introducing flexibility through parameterization based on Rényi divergence \cite{mironov:renyi}. Compared with ($\epsilon, \delta$)-DP, RDP provides robust probabilistic privacy guarantees without the risk of complete privacy breaches that are permissible under the ($\epsilon, \delta$)-DP framework, which allows a $\delta$ probability of total information disclosure \cite{mironov:renyi}. RDP's guarantees are contingent on outcome probabilities, thus prohibiting absolute privacy violations and preserving uncertainty even under weak parameters  \cite{mironov:renyi}.

However, this dependency necessitates complex baseline risk assessments, especially challenging in large or dynamic datasets where probabilities are elusive. Despite its theoretical benefits, RDP's integration into cryptographic frameworks can be cumbersome due to the requisite precise assessments and their complex incorporation with cryptographic arguments. This complexity makes RDP less suited for demonstrating the feasibility of privacy-preserving computations in practical applications, especially those involving co-processing arrangements where trust and verifiability are paramount. In contrast, DP-OLS offers a simpler, direct method by embedding privacy controls within the regression algorithm \cite{shef:ols}, facilitating easier application, verification, and validation in cryptographic contexts, aligning effectively with robust, privacy-preserving research objectives.

\subsection{($\epsilon$, $0$) Differentially Private Regressions Over Linear Subspaces}

\noindent A linear regression is a simple form of machine-learning algorithm which attempts to find a line passing through a set of points such that the distance, referred to further as the \textit{error} of each predicted value with respect to the line is minimized.

For a domain of values $\textbf{x}=(x_1...x_n)^T$ mapped to a range of features $\textbf{y}=(y_1...x_n)^T$:
\[
\tilde{x} = \frac{1}{n} \sum^n_{i=1}x_i, 
\tilde{y} = \frac{1}{n} \sum^n_{i=1}y_i
\]
\[\text{with ncov}(x,y)= \left \langle \textbf{x}-x_1, \textbf{y}-y_1 \right \rangle\]
\[\text{and nvar}(x,y)= \left \langle \textbf{x}-x_1, \textbf{x}-x_1 \right \rangle= n \cdot \text{var}(\textbf{x})
\]
The noisy linear regression, in matrix notation, is defined as $Y = \alpha \cdot \textbf{x} + \beta + \textbf{e}$:
\[
\begin{bmatrix}
    y_1 \\
    y_2 \\
    . \\
    . \\
    . \\
    y_n \\
\end{bmatrix} 
=
\alpha
\begin{bmatrix}
    1 & x_1 \\
    1 & x_2 \\
    . & . \\
    . & . \\
    . & . \\
    1 & x_n \\
\end{bmatrix}
+
\begin{bmatrix}
    \beta_1 \\
    \beta_2 \\
    . \\
    . \\
    . \\
    \beta_n \\
\end{bmatrix}
+\textbf{e}
\]

\noindent For a noise distribution $F_e(0,\sigma^2_e)$ and cost function "mean squared error" defined as:

\[\mathscr{L(\theta)} = \frac{1}{n} \sum^n_{i=0} (y_i-(\alpha x_i+ \beta_i))^2 \]

\subsection{Differential Privacy With NoisyStats}

\noindent We select the compact ($\epsilon,0$)-DP NoisyStats \cite{alabi:mcmil:sar:smi:vad} algorithm as a means of introducing differential privacy which perturbs values in the $\beta$ vector with noise drawn from the Laplace mechanism. As specified in \cite{alabi:mcmil:sar:smi:vad:reg}, we fail if the denominators of noisy versions of $ncov$ and $cov$ become less than or equal to zero.

\begin{algorithm}[H]%
\caption{NoisyStats: ($\epsilon=2$, 0)-DP }\label{alg:cap}
\begin{algorithmic}
    \State \textbf{Data: $\left \{ \left (  x_i, y_i\right ) \right \}^n_{i=1}\in \left ( \left [ 0,1 \right ] \times\left [ 0,1 \right ] \right )^n$}
    \State \textbf{Privacy Params: $\epsilon$}
    \State \textbf{Hyperparams: }\text{none}\\
    Define $\Delta_1 = \Delta_2= (1-1/n)$\\
    Sample $L_1\sim $ Lap(0,3$\Delta_1$/$\epsilon$)\\
    Sample $L_2\sim $ Lap(0,3$\Delta_2$/$\epsilon$)
    \If{\textit{nvar(\textbf{x})$+L_2>0$}}
        \State $\tilde{\alpha} = \frac{ncov(\textbf{x},\textbf{y})+L_1}{nvar(\textbf{x})+L_2}$
        \State $\Delta_3 = 1/n \cdot (1+|\tilde{\alpha}|)$
        \State $\text{Sample } L_3\sim $ Lap(0,3$\Delta_3$/$\epsilon$)
        \State $\tilde{\beta} = (\tilde{y}-\tilde{\alpha}\tilde{x}) +L_3$
        \State \textbf{return} $\tilde{\alpha}+ \tilde{\beta}$
    \Else
        \State \textbf{return} $\bot$
    \EndIf
\end{algorithmic}
\end{algorithm}
\noindent Algorithm 1 produces $\beta$, which is now ($\epsilon,0$) differentially-private. This technique for private regression modeling is well-studied and common in relevant literature \cite{alabi:mcmil:sar:smi:vad:reg}, \cite{base:dp}. The novelty of our approach lies in the pairing of this scheme with a cryptographic computational attestation of integrity, such that our protocol definition from section \ref{sec:protocol} is satisfied. Perhaps most importantly, however, is how we shift the workload of the computation from the verifier to the prover, which is presented further.

\section{Obtaining Computational Integrity Statements}
\label{sec:zkprotocols}

\noindent There are two major approaches towards producing a protocol that satisfies the requirements from section \ref{sec:protocol}. The first means of achieving a computational attestation of integrity for a differentially-private regression resembles that of constructing an application-specific integrated circuit (ASIC) which describes an algorithm or computation directly as a circuit using standardized constraint systems (R1CS, AIR, etc).  In a fashion not entirely dissimilar from that of the early days of computing, a computation to be proven is encoded as a simulation of a circuit, which is defined over gates and wires. This "ASIC approach" is thus far common in ZK literature as a consequence of the field of study being relatively young, and as a result, few robust general-purpose ZK frameworks and languages exist at the time of this writing, and even fewer for a STARK-based approach.

Recent technological breakthroughs \cite{risczero:about} have led to the discovery that entire instruction set architectures (ISAs) can be encoded in such a circuit. It is then possible to use this construction to simulate a small virtual machine, which is executed as a provable circuit. An argument of valid computation arising from such a system is structured to show the correct transition of values stored in the registers of this machine given a particular set of instructions. This design is referred to as a zero-knowledge virtual machine (ZKVM). A ZKVM can be built on top of a wide variety of argument systems. We select the RISC-Zero virtual machine \cite{risczero:about}, which constructs an argument system that satisfies the protocol described in section \ref{sec:protocol}. There are two key features of this ZKVM which we leverage in our result:

\begin{itemize}
    \item The RISC-Zero ZKVM accepts arbitrary Rust code as input, leading to a substantial reduction in programming complexity and iteration time as far as algorithm description is concerned. This approach also facilitates the development of a richer set of applications that can be expressed due to the relative ease of using high-level languages to describe a computation to be proven. 

    \item As a virtual representation of a small RISC-V machine, the RISC-Zero ZKVM is compatible with essentially any Rust crate which can be targeted to the RISC-V ISA. Out-of-the-box, this platform is immediately usable by many existing libraries, types and software in the Rust ecosystem.
    
\end{itemize}

Certain functions and algorithms will exhibit performance characteristics which may vary substantially depending on the hardware in which they are executed. Modern computing hardware typically includes specialized circuits designed for targeted applications (such as cryptographic hashing and elliptic curve operations), and these sub-circuits exhibit runtime performance several orders of magnitude above what a central processing unit (CPU) may provide. The ZKVM is a software simulation of a RISC-V computer, so it is executed using CPU hardware. We remark that executing the simulation in this way incurs substantial runtime and performance overhead, as we would expect for any software simulation of a circuit. The RISC-Zero ZKVM is packaged with an optional CUDA backend which can leverage GPU hardware for the complex proving machinery in the background. We observe a substantial reduction in runtime when using this feature, even on our relatively small and modest laptop GPU.

\section{Our Results}
\label{sec:observations}

\noindent This section defines the experimental setup of the study and the results we obtain. We run our provable differentially-private regression on a single debian-based laptop with a 13th Gen Intel Core™ i9-13900H CPU, 32 GB of RAM, and an NVIDIA GeForce RTX™ 4070 Laptop GPU with 8gb of available VRAM. We observe that this relatively low-cost and modest hardware is more than sufficient to carry out our experiments and demonstrates how protocols meeting the attributes defined in Section \ref{sec:protocol} are not necessarily isolated to high-power and high-cost cloud-based platforms.

For our experiments, we use the "Kaggle Healthcare Dataset" \cite{pras:health} which contains 50,000 synthetic records of patients admitted to hospital care, several factors regarding their health, and the resulting insurance amount billed for the visit. For a linear-regression over $(x, y)$ pairs, we remove all columns except for age and insurance cost, and learn a function that approximates the relationship between the two.

\subsection{Training and Experiment Design}

\noindent We learn two hypotheses: the first is based on a simple "ordinary least squares (OLS)" error estimator, and the second is a differentially private version of OLS (DP-OLS). We observe that the error of the DP-OLS hypothesis initially diverges significantly from the OLS hypothesis when trained on few data samples. This is an expected result given how the data is perturbed in the DP-OLS training, and with few samples to learn from, the noisy hypothesis should display a strong divergence from the OLS estimator.

Metrics that we are interested in include "Proof time versus Dataset Size", "Verification Time vs Dataset Size", and "DP vs OLS Model Accuracy". We conduct the linear-regression training over the entire dataset using both the CPU and the GPU in our laptop. Due to the current limitations of the ZKVM and our GPU hardware, we must prove the dataset in batches when using the CUDA feature of the ZKVM. The GPU has limited VRAM which is quickly exceeded during proof generation. This limitation does not affect the overall runtime of the experiment, rather it just causes proofs over large datasets to be carried out in separate runs of the ZKVM.

The ZKVM immediately supports all primitive rust types in a "\text{no\_std}" environment. The calculations involved in the Laplace Mechanism are carried out over the rationals, so initially we performed all of our arithmetic using the built-in f32 types which are IEEE floating-point decimals. This type lends relative ease towards rational arithmetic, however we find that we can prove a far greater number of samples per batch when using fixed-point arithmetic with the GPU before exceeding the available VRAM. We can prove up to 1400 data points per batch with fixed-point representation, vs only 175 samples per batch using floating-point representation. The charts below illustrate the measurements that we collect during our study.

\begin{figure}[H]
    \centering
    \begin{minipage}{0.48\textwidth}
        \centering
        \begin{tikzpicture}
        \begin{axis}[
            ybar,
            enlarge x limits=0.25,
            width=\textwidth,
            height=5cm,
            ylabel={Total Proving Time (seconds)},
            xlabel={Hardware \& Number System},
            symbolic x coords={CPU-FX, GPU-FL, GPU-FX},
            xtick=data,
            nodes near coords,
            ymin=0, ymax=1800,
            bar width=30pt,
            title={Proving Time vs. Arithmetic Datatype},
            legend style={at={(0.5,-0.20)}, anchor=north, legend columns=-1}
        ]
        \addplot coordinates {(CPU-FX,1560) (GPU-FL,858) (GPU-FX,360)};
        \end{axis}
        \end{tikzpicture}
        \caption{CPU fixed-point vs. GPU floating-point vs. GPU fixed-point proving time over the entire dataset of 50,000 samples.}
        \label{fig:bar_graph}
    \end{minipage}\hfill
    \begin{minipage}{0.48\textwidth}
        \centering
        \begin{tikzpicture}
        \begin{axis}[
            title={Runtime vs. Dataset Size},
            xlabel={Dataset Size},
            ylabel={Runtime (seconds)},
            xmin=0, xmax=5000,
            ymin=0, ymax=3000,
            xtick={0, 1000, 2000, 3000, 4000, 5000},
            xticklabels={0, 10k, 20k, 30k, 40k, 50k},
            ytick={0, 500, 1000, 1500, 2000, 2500, 3000},
            yticklabels={0, 0.5k, 1k, 1.5k, 2k, 2.5k, 3k},
            legend pos=north west,
            grid style=dashed,
            width=\textwidth,
            height=5cm
        ]
        \addplot[color=blue, mark=square] coordinates {(0, 0) (500,36) (1000,72) (1500,108) (2000,144) (2500,180) (3000,216) (3500,252) (4000,288) (4500,324) (5000,360)};
        \addplot[color=red, mark=square] coordinates {(0, 0) (500,85.8) (1000,171.6) (1500,257.4) (2000,343.2) (2500,429) (3000,514.8) (3500,600.6) (4000,686.4) (4500,772.2) (5000,858)};
        \addplot[color=purple, mark=triangle] coordinates {(0, 0) (500,156) (1000,312) (1500,468) (2000,624) (2500,780) (3000,936) (3500,1092) (4000,1248) (4500,1404) (5000,1560)};
        \legend{GPU-Fixed, GPU-Float, CPU-Fixed}
        \end{axis}
        \end{tikzpicture}
        \caption{Runtime vs. dataset size over different data types. We measure the performance of floating-point vs. fixed-point decimals on both the GPU and the CPU hardware.}
        \label{fig:runtime_graph}
    \end{minipage}
\end{figure}

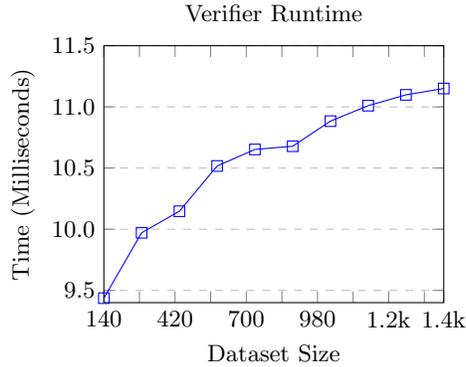
\begin{figure}[H]
    \centering
    \begin{tikzpicture}
    \begin{axis}[
        title={Verifier Runtime},
        xlabel={Dataset Size},
        ylabel={Time (Milliseconds)},
        xmin=1.4, xmax=14, 
        ymin=9.4, ymax=11.5,
        xtick={1.4, 2.72, 4.04, 5.36, 6.68, 8, 9.32, 10.64, 11.96, 13.28, 14}, 
        xticklabels={140, , 420, , 700, , 980, , 1.2k, , 1.4k}, 
        ytick={9.5, 10.0, 10.5, 11.0, 11.5},
        yticklabels={9.5, 10.0, 10.5, 11.0, 11.5},
        legend pos=north west,
        ymajorgrids=true,
        grid style=dashed,
        width=0.5\textwidth, 
        height=5cm
    ]
    \addplot[
        color=blue,
        mark=square,
    ]
    coordinates {
        (1.4,9.436543)(2.8,9.971104)(4.2,10.147017)(5.6,10.51707)(7,10.652056)
        (8.4,10.678342)(9.8,10.883507)(11.2,11.009815)(12.6,11.09831)(14,11.150112)
    };
    \end{axis}
    \end{tikzpicture}
    \caption{Verifier runtime over a single GPU-Float batch divided into 10 minibatches}
    \label{fig:verifier_runtime}
\end{figure}

\noindent Figure \ref{fig:bar_graph} represents the observed runtimes during the experiment. We record the runtimes differently depending on the underlying hardware and number system in use for the proof system. The first and largest runtime measurement, CPU-FX, is the total time taken in seconds to prove the DP-training of a linear regression over the entire dataset of 50,000 samples using a fixed-point number system on the CPU. The next measured runtime, GPU-FL, measures the total time taken to prove the DP-training over the entire dataset using the GPU hardware and the floating-point number system. Lastly, GPU-FX is the time taken for the same task using the GPU and a fixed-point number system. The differences in runtime between hardware and number systems are illustrated clearly in the chart. We find that a GPU with a fixed-point number representation performs the best in terms of runtimes out of all measured configurations. We do not collect a CPU-based floating-point representation in these experiments. 

Figure \ref{fig:runtime_graph} conducts the same measurements as in figure \ref{fig:bar_graph}, but we collect the measured proving time as a function of dataset size. This is done to expressly illustrate the linear growth rate of the prover as the dataset size increases. The graph suggests that the prover is meeting the protocol specified in \ref{sec:protocol}. For GPU-based measurements, the limitation of computation size becomes a factor, and thus we divide the dataset into batches of 1400 elements and collect the total runtime by aggregating each batch runtime up to the dataset size.

Lastly, we measure and record the runtime of the verifier in figure \ref{fig:verifier_runtime}. We observe the measurements of the verifier runtime to increase in a roughly logarithmic fashion with respect to the dataset size. This measurement was observed to be substantially less consistent over multiple runs, in contrast to the prover runtime. Nonetheless, the relationship between runtime and dataset size for the verifier appears to be logarithmic, which is the expected result and further validates that the ZKVM used for this task conforms to the protocol specifications in \ref{sec:protocol}.

The verifier is described as a small and relatively simple computation. As such, it does not require specialized hardware or optimized number systems. We observe that the verifier is convinced of the authenticity of the prover's CI statement for the task of proving the differentially-private training of the linear regression over 50,000 samples in approximately 0.17 seconds, regardless of the number system or hardware used by the prover. We observe a very slow, logarithmic growth rate of the verifier over increasing dataset sizes, as illustrated above. This observation highlights the utility of this argument system; the verifier can obtain complex and expensive results with relative ease by working in cryptographic concert with a powerful prover.

\subsection{Loading the Dataset}

\noindent The RISC-Zero ZKVM is packaged with a simulation of a standard input and output, established as a channel between the prover and verifier (this is entirely distinct from the cryptographic proof channel of the STARK protocol underlying the argument system). This channel allows the verifier to send data to the prover. We initially used this channel to send our dataset into the prover, however we observed during early experiments that this step had a significant and detrimental effect on the prover performance. The verifier sends messages to the prover as bytes by serializing any message they wish the prover to receive. The prover then safely deserializes the message by checking that the bytes conform to the expected data. 

This deserialization step is included in the CI statement, and is observed to incur a significant degradation in prover performance. We mitigate this issue by forgoing use of the channel entirely, and directly embed the bytes into the binary that we send to the prover. Instead of performing a costly deserialization, the prover simply conducts a pointer-cast to the embedded bytes \cite{chen:l2iv}. In our experiments, this results in a dramatic reduction from approximately 232 million cycles to roughly \textit{640 cycles} to load the dataset into the RAM of the ZKVM. Figure \ref{fig:data_handling} illustrates our approach and highlights the savings in computation cycles when choosing this method.

\usetikzlibrary{positioning, shapes.geometric, arrows}

\begin{figure}[H]
\centering

\tikzset{
    block center/.style ={rectangle, draw=black, thick, fill=white,
                          text width=8em, text centered,
                          minimum height=2em},
    line/.style ={draw, thick, ->, shorten >=2pt},
    blueline/.style ={draw, thick, ->, shorten >=2pt, blue}
}

\begin{tikzpicture}[node distance=1cm and 1cm, auto, thick, >=latex]

    \node[block center] (verifier) {Verifier};
    \node[block center, right=of verifier] (channel) {Communication Channel};
    \node[block center, right=of channel] (prover) {Prover};
    \node[block center, below=of channel] (deserialization) {Deserialization};
    \node[block center, below=of deserialization] (typesafety) {Type-Safety Check};
    \node[block center, below=of typesafety] (pointer) {Dataset Loaded};

    \path[line] (verifier) -- (channel);
    \path[line] (prover) -- (channel);
    \path[line] (prover) -- (deserialization);
    \path[line] (deserialization) -- (typesafety);
    \path[line] (typesafety) -- (pointer) node[midway, left] {232m Cycles};
    \path[blueline] (prover.south) |- (pointer) node[near start, left] {640 Cycles};

\end{tikzpicture}

\caption{Optimized Data-Handling Path in the RISC-Zero ZKVM}
\label{fig:data_handling}
\end{figure}
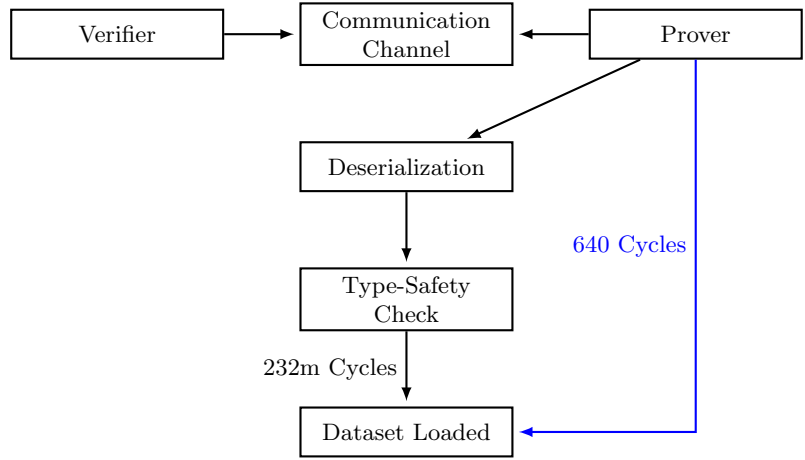

\noindent This technique requires the use of "unsafe Rust", which in some settings may impact an application's security. In our case, we assume that the verifier competently constructs the binary with the embedded bytes and has performed due-diligence to ensure the memory-safety of the computation that they send to the prover. Further, because the size of our dataset is known in advance, it is represented in Rust as an array in contiguous memory, rather than being stored in non-contiguous heap-allocated memory. The only manner in which a prover can generate a valid CI statement is by executing the binary faithfully. They do not choose array bounds or make any modifications to the program to be executed without causing the CI statement to become invalid, with high probability. Thus, as long as the verifier has ensured that the memory access patterns of the binary are safe and correct, this technique does not pose a risk to the correct execution of the program.

With the bytes embedded into the ZKVM in this fashion, communication between the prover and verifier is simplified further. All the verifier needs to do is simply distribute the compiled binaries to any party who is willing to execute them. There is no longer a need to communicate with the prover after the protocol has been established, outside of receiving the CI statement attesting that the binary was executed successfully.

\subsection{Validation}

\begin{table}[H]
\centering
\captionsetup{skip=4pt} 
\caption{Standard vs. Noisy Regression Models}
\label{tab:model_accuracy}
\setlength{\tabcolsep}{8pt} 
\begin{tabular}{@{}lccc@{}} 
\toprule
Metric & OLS & DP-OLS & Delta \\ \midrule
Slope std. Error & 0.01636 & 0.01634 & 0.000018 \\
Intercept std. Error & 0.90285 & 0.90285 & $< 0.00001$ \\
Mean Absolute Error & 12205.307 & 12205.302 & 0.005252 \\
\bottomrule
\end{tabular}
\end{table}

\noindent Table \ref{tab:model_accuracy} measures the overall accuracy of the predictions made by both the DP and the non-DP estimators. The difference between mean absolute error for both regressions is low after training over 50,000 samples, indicating that the DP regression makes predictions almost identical to that of the non-DP regression. These results both strongly suggest that despite the noisy training data, the DP estimator converges to almost exactly the same model as the non-DP estimator.

\begin{table}[H]
\centering
\captionsetup{skip=4pt} 
\caption{Progression of training accuracy over multiple iterations}
\label{tab:model_data}
\setlength{\tabcolsep}{7pt} 
\begin{tabular}{cccccc} 
\toprule
Iteration & Age Label & Cost Label & Slope & Intercept & Noisy Intercept \\
\midrule
1 & 25.81 & 31664.69 & 2647.30 & -5397.57 & -5396.66 \\
2 & 72.82 & 30000.73 & 491.68 & 23344.14 & 23345.05 \\
3 & 79.36 & 25865.49 & -1111.96 & 45260.56 & 45261.47 \\
4 & 48.65 & 27202.88 & -942.82 & 42892.53 & 42893.44 \\
5 & 26.28 & 26494.02 & -861.07 & 41720.79 & 41721.70 \\
6 & 59.63 & 25762.27 & -818.02 & 41089.36 & 41090.27 \\
7 & 72.17 & 25153.07 & -782.68 & 40559.26 & 40560.17 \\
8 & 56.42 & 24992.15 & -724.74 & 39670.81 & 39671.72 \\
9 & 74.09 & 25296.88 & -639.09 & 38328.95 & 38329.86 \\
10 & 24.75 & 26023.56 & -530.29 & 36588.17 & 36589.08 \\
\bottomrule
\end{tabular}
\end{table}

\noindent Table \ref{tab:model_data} measures the values for the model slope and intercept as training is carried out over the dataset. As larger volumes of samples are trained over, we observe both estimators converging to roughly the same hypothesis, which is again the expected result. This indicates that DP-OLS, despite the noise perturbation, is successfully learning approximately the same hypothesis as the OLS regression. If both estimators eventually settled on drastically different slopes and intercepts, we would take this as an indication that the parameters for DP were either incorrect or set too aggressively.

\subsection{Scaling Up}

\noindent The limitation of batch-size within the ZKVM is only a limitation on a single ZKVM running on a single GPU. We observe a means of leveraging the batched approach within a setting in which proving time could be further dramatically reduced, given access to \textit{multiple} machines. In this setting, proving time becomes a function of each new machine joining the system, up to an optimal number of nodes. When considering the overhead of instantiating the ZKVM and generating the proofs, we find that 1400 samples is the optimal batch size for each ZKVM to prove on our hardware setup. If 50,000 samples are evenly divided across a network of 36 single GPUs (assuming each GPU is the same as that used in our experiment), the entire dataset could be proven \textit{in less than 10 seconds}. The Ethereum blockchain network is recorded \cite{mining:model} as having a compute power of 603,000 GHz on average at any given time in 2021. Our GPU has a clock speed of 1.61 GHz \cite{note:gpu}, suggesting that in 2021, there existed a distributed network of approximately 374,534 available "GPU units" available for compute at any given time.

Lastly, the task of aggregating the individually computed models into a single regression remains. This step requires $O(c)$ work with respect to the number of nodes in the system and is carried out by the verifier. It is assumed at this point that the verifier, as in our experiments, has already carried out verification of each received proof and can trust with high probability the correctness of each model it has received from a node. Each linear regression model is represented by the equation:
\[ y = \beta_1 x + \beta_0 \]
where \( \beta_1 \) is the slope and \( \beta_0 \) is the intercept. To combine models from multiple parties, the slopes (\( \beta_1 \)) and intercepts (\( \beta_0 \)) of these models are averaged.

1. Averaging Slopes:
   \[ \overline{\beta_1} = \frac{1}{n} \sum_{i=1}^n \beta_{1i} \]
   where \( n \) is the number of parties (or models), and \( \beta_{1i} \) is the slope of the \( i \)-th model.

2. Averaging Intercepts:
   \[ \overline{\beta_0} = \frac{1}{n} \sum_{i=1}^n \beta_{0i} \]
   where \( \beta_{0i} \) is the intercept of the \( i \)-th model. The combined model is then given by:
\[ y = \overline{\beta_1} x + \overline{\beta_0} \]

\section{Comparison To Other Works}
\label{sec:comparison}

\noindent We assess the literature surrounding the topic of provable-DP to be relatively sparse. The closest result to ours is found in \textbf{confidential-DP} \cite{sham:confdp}. To the best of our understanding, this work is derived from a zero-knowledge construction contained within the EMP toolkit \cite{weik:emp}, itself based on an argument system referred to as "Wolverine" \cite{weng:wolv}. To the best of our knowledge, confidential-DP does not report a verifier runtime, and so we are left to assess the complexity of the underlying argument system in use. We compare our protocol to confidential-DP in Table \ref{tab:protocols}.

It is implied by \cite{weng:wolv} that confidential-DP obtains an $O(n)$ prover and an $O(n)$ verifier. By contrast, our work obtains an $O(quasi(n))$ prover and an $O(polylog(n))$ verifier. confidential-DP is an interactive argument system, requiring the proving and verifying parties to be online and communicating during the execution of the protocol. Our result is non-interactive, the prover and verifier only exchange a single message to instantiate the protocol, and a single message from the prover is sent containing the result and proof of correctness. Both results are zero-knowledge, in that the verifying party need not be the same who initiated the protocol, and the argument of integrity is zero-knowledge and reveals nothing except that the argument is correct.

\begin{table}[h]
\centering
\captionsetup{skip=4pt} 
\caption{Asymptotic comparison of this work and confidential-DP}
\label{tab:protocols}
\footnotesize 
\begin{tabular}{
  |>{\centering\arraybackslash}m{1.75cm}
  |>{\centering\arraybackslash}m{1.75cm}
  |>{\centering\arraybackslash}m{1.5cm}
  |>{\centering\arraybackslash}m{1.5cm}
  |>{\centering\arraybackslash}m{1.5cm}
  |>{\centering\arraybackslash}m{1.5cm}|
}
\hline
\textbf{\rule{0pt}{2.5ex}Protocol} & \textbf{\rule{0pt}{2.5ex}DP} & \textbf{\rule{0pt}{2.5ex}Model} & \textbf{\rule{0pt}{2.5ex}Prover} & \textbf{\rule{0pt}{2.5ex}Verifier} & \textbf{\rule{0pt}{2.5ex}Online} \\ \hline
This work & NoisyStats, ($\epsilon = 2$, 0) & Linear & $O(n)$ & $O(\log n)$ & No \\ \hline
confidential-DP & DP-SGD, $\epsilon=0.55$, $\delta = 10^{-5}$ & Logistic & $O(n)$ & $O(n)$ & Yes \\ \hline
\end{tabular}
\end{table}

\noindent Other observed differences between results are that confidential-DP classifies images, while our model predicts labels from data points represented as pairs of single numbers. Confidential-DP reports a total proving time of 100 hours over the CIFAR10/MNIST dataset with 60,000 samples, while we observe a best-case 6 minute proving time over our Kaggle Healthcare dataset with 50,000 samples. We remark however that runtime in this context is not a meaningful comparison of performance between the two results, since the tasks are performed over different data with different model architectures. Our aforementioned analysis of verifier complexity stands as the major difference between this result and \cite{sham:confdp}.

Lastly, confidential-DP proves only the DP portion of their training. They first extract features from the training data with a deep network, then train on those features with a DP-logistic regression. This is done because describing an entire deep network (particularly the computation graphs and gradients for back-propagation) as a ZK-circuit remains a highly non-trivial and contrived task. Our work argues the integrity of the entire training process, however our training regimen is a comparatively simple model consisting of a compact number of operations that are readily programmed into the ZKVM.

\section{Conclusion}
\label{sec:conclusion}

 \noindent This work shows a robust and efficient means of cryptographically proving and verifying the correct execution of an agreed-upon machine-learning training process. We achieve record-breaking proving times for a differentially-private, simple linear-regression over a large dataset, and we show how to effectively optimize large data transfers into zero-knowledge virtual machines. We leverage modest but powerful graphics hardware to accelerate proving performance orders of magnitude over a standard, CPU-based approach. We also describe a means of establishing a network of machines to prove sections of the data in parallel, which can lead to further dramatic reductions in proving time for cryptographically-secure machine-learning training. 
 
 We improve over the state-of-the-art verifier performance in this research domain from an $O(n)$ verifier to a $O(log(n))$ verifier, showing how a differentially-private regression can be securely obtained with high confidence from an untrusted third-party in a fraction of a second. Lastly, we show how the use of fixed-point arithmetic over IEEE floating point numbers leads to a dramatic improvement in prover runtimes. The use of integer-based arithmetic in this fashion lays a foundation for realizing more complex operations in the future, such as gradient computation for back-propagation training. 
 
 We believe our design results in an attestation of integrity which shows that machine-learning algorithms, (albeit simple algorithms, for the time-being) are well-suited to this particular form of non-interactive argument system, demonstrating a practical application of introducing an "asymmetry" into a heavy computation in order to leverage powerful hardware which may exist in a different physical or temporal location. Our result shows how a verifying party can apply this framework to efficiently obtain an irrefutably correct machine-learning model from an untrusted but powerful outside source.

Argument systems of this variety facilitate what seems at first to be a counter-intuitive, yet intriguing result; by working in concert with a prover, the verifier has learned the exact same machine-learning model, but appears to have done so with only $O(polylog(n))$ work with respect to the dataset size \cite{gold:roth:shaf:yehu}. As the dependence on outsourced ML hardware continues to grow, we anticipate the need for secure "co-processing" solutions of this nature to expand in kind. We believe this work shows that state-of-the-art ZKVM constructions are uniquely equipped to play an important role in the growth of privacy-preserving machine-learning.

We believe that this research constitutes a vital advancement in the development of completely private end-to-end machine-learning, in which distinct and distrustful parties (model operators and consumers) may interact with each other in complete privacy. During the inference phase of this research, the model itself and the data to be classified are revealed out of necessity. This could hypothetically be remedied by recent advancements in the field of fully-homomorphic encryption. \cite{zama:ml} demonstrates the practical feasibility of conducting regressions over encrypted data, while \cite{chen:l2iv} leverages the RISC-Zero ZKVM to prove fully-homomorphic computations over encrypted ciphertexts. Future work targets the understanding of the costs of fully homomorphic and provable differentially-private regressions and classifications using the ZKVM, along with other frameworks and tools. We include all experiments and the entire construction from this work in a GitHub repository \cite{ray:capy}. The code is readily executed with ease assuming the proper hardware configuration.
\\
%
%


\begin{thebibliography}{6}
%

\bibitem {deloitte:genai}
Deloitte: Generative Artificial Intelligence.
Available online: \url{https://www2.deloitte.com/us/en/pages/consulting/articles/generative-artificial-intelligence.html}
[Accessed 20-09-2023]

\bibitem {gold:roth:shaf:yehu}
Goldwasser, S., Rothblum, G., Shafer, J., Yehudayoff, A.: ECCC - TR20-058.
Available online: \url{https://eccc.weizmann.ac.il/report/2020/058/}
[Accessed 20-09-2023]

\bibitem {ben:chie:spoo}
Ben-Sasson, E., Chiesa, A., Spooner, N.: Interactive Oracle Proofs.
Cryptology ePrint Archive, Paper 2016/116, 2016.
\url{https://eprint.iacr.org/2016/116}

\bibitem {ben:ben:hor:ria}
Ben-Sasson, E., Bentov, I., Horesh, Y., Riabzev, M.: Scalable, transparent, and post-quantum secure computational integrity.
Cryptology ePrint Archive, Paper 2018/046, 2018.
\url{https://eprint.iacr.org/2018/046}

\bibitem {alabi:mcmil:sar:smi:vad}
Alabi, D., McMillan, A., Sarathy, J., Smith, A., Vadhan, S.: Differentially Private Simple Linear Regression.
arXiv preprint arXiv:2007.05157, 2020.
\url{https://arxiv.org/abs/2007.05157}

\bibitem {lee:kif}
Lee, J., Kifer, D.: Scaling up Differentially Private Deep Learning with Fast Per-Example Gradient Clipping.
arXiv preprint arXiv:2009.03106, 2020.
\url{https://arxiv.org/abs/2009.03106}

\bibitem {abadi:deep}
Abadi, M., Chu, A., Goodfellow, I., McMahan, H. B., Mironov, I., Talwar, K., Zhang, L.: Deep Learning with Differential Privacy.
In: Proceedings of the 2016 ACM SIGSAC Conference on Computer and Communications Security, pp. 308–318. Association for Computing Machinery, New York, NY, USA, 2016.
\url{https://doi.org/10.1145/2976749.2978318}

\bibitem {goog:ai:dp}
Applying differential privacy to large scale image classification.
Google AI Blog, 2022, Feb.
\url{https://ai.googleblog.com/2022/02/applying-differential-privacy-to-large.html}

\bibitem {dwo:rot}
Dwork, C., Roth, A.: The algorithmic foundations of Differential Privacy.
Foundations and Trends® in Theoretical Computer Science, vol. 9, no. 3-4, pp. 211–407. 2013.
\url{https://doi.org/10.1561/0400000042}

\bibitem {fath:noise}
Fathima, S.: Differential Privacy-noise adding mechanisms.
Medium, Becoming Human: Artificial Intelligence Magazine, 2020, Oct.
\url{https://becominghuman.ai/differential-privacy-noise-adding-mechanisms-ede242dcbb2e}

\bibitem {wiki:laplace}
Laplace distribution.
Wikipedia, Wikimedia Foundation, 2022, May.
\url{https://en.wikipedia.org/wiki/Laplace_distribution}

\bibitem {stei:ull}
Steinke, T., Ullman, J.: Between Pure and Approximate Differential Privacy.
arXiv preprint arXiv:1501.06095, 2015.
\url{https://arxiv.org/abs/1501.06095}

\bibitem {dwo:mcs:nis:smi}
Dwork, C., McSherry, F., Nissim, K., Smith, A.: Calibrating Noise to Sensitivity in Private Data Analysis.
In: Theory of Cryptography, Third Theory of Cryptography Conference, TCC 2006, Lecture Notes in Computer Science, vol. 3876, pp. 265-284. Springer, 2006.
\url{https://iacr.org/archive/tcc2006/38760266/38760266.pdf}

\bibitem {das:mish}
Das, S., Mishra, S.: Advances in Differential Privacy and Differentially Private Machine Learning.
In: Springer Tracts in Electrical and Electronics Engineering. Springer Nature Singapore, pp. 147–188, 2024.
\url{http://dx.doi.org/10.1007/978-981-97-0407-1_7}

\bibitem {bun:ull:vad}
Bun, M., Ullman, J., Vadhan, S.: Fingerprinting Codes and the Price of Approximate Differential Privacy.
arXiv preprint arXiv:1311.3158, 2018.
\url{https://arxiv.org/abs/1311.3158}

\bibitem {wang:lei:fien}
Wang, Y.-X., Lei, J., Fienberg, S. E.: Learning with Differential Privacy: Stability, Learnability and the Sufficiency and Necessity of ERM Principle.
arXiv preprint arXiv:1502.06309, 2016.
\url{https://arxiv.org/abs/1502.06309}

\bibitem {mironov:renyi}
Mironov, I.: Rényi Differential Privacy.
In: 2017 IEEE 30th Computer Security Foundations Symposium (CSF), IEEE, 2017, Aug.
\url{http://dx.doi.org/10.1109/CSF.2017.11}

\bibitem {shef:ols}
Sheffet, O.: Differentially Private Ordinary Least Squares.
arXiv preprint arXiv:1507.02482, 2017.
\url{https://arxiv.org/abs/1507.02482}

\bibitem {alabi:mcmil:sar:smi:vad:reg}
Amin, K., Joseph, M., Ribero, M., Vassilvitskii, S.: Easy Differentially Private Linear Regression. arXiv preprint arXiv:2208.07353, 2023.
\url{https://arxiv.org/abs/2208.07353}

\bibitem {base:dp}
Implement differential privacy with tensorflow privacy: responsible AI toolkit.
TensorFlow. Available online: \url{https://www.tensorflow.org/responsible_ai/privacy/tutorials/classification_privacy}

\bibitem {risczero:about}
RiscZero: About.
Available online: \url{https://www.risczero.com/about}
[Accessed 20-09-2023]

\bibitem {pras:health}
Prasad, S.: Healthcare Dataset.
Available online: \url{https://www.kaggle.com/datasets/prasad22/healthcare-dataset}, 2022.
[Accessed 12-06-2024]

\bibitem {chen:l2iv}
Chen, W., Research Partner, L2IV (@weikengchen): Tech Deep Dive: Verifying FHE in RISC Zero, Part I.
Available online: \url{https://l2ivresearch.substack.com/p/tech-deep-dive-verifying-fhe-in-risc}, 2024.
[Accessed 18-04-2024]

\bibitem {mining:model}
Mining Model Spreadsheet: Mining Model Spreadsheet.
Available online: \url{https://docs.google.com/spreadsheets/d/138M4R1-_zS-OLBsl2VJeN_anfTSCRCFc6EguYUVG-yA/edit#gid=1339763553}
[Accessed 20-09-2023]

\bibitem {note:gpu}
NotebookCheck: NotebookCheck.
Available online: \url{https://www.notebookcheck.net/NVIDIA-GeForce-RTX-4070-Laptop-GPU-Benchmarks-and-Specs.675690.0.html}, 2024.
[Accessed 18-04-2024]

\bibitem {sham:confdp}
Shamsabadi1 et al: Confidential-DPproof: CONFIDENTIAL PROOF OF DIFFERENTIALLY PRIVATE TRAINING.
Available online: \url{https://openreview.net/pdf?id=PQY2v6VtGe}, 2024.
[Accessed 18-04-2024]

\bibitem {weik:emp}
Weikeng et al: emptoolkit.
Available online: \url{https://github.com/emp-toolkit/emp-zk}, 2023.
[Accessed 18-04-2024]

\bibitem {weng:wolv}
Weng, C. et al: Wolverine: Fast, Scalable, and Communication-Efficient Zero-Knowledge Proofs for Boolean and Arithmetic Circuits.
Available online: \url{https://eprint.iacr.org/2020/925.pdf}, 2020.
[Accessed 18-04-2024]

\bibitem {zama:ml}
Zama: Concrete ml.
GitHub, 2024.
\url{https://github.com/zama-ai/concrete-ml}
[Accessed 20-09-2023]

\bibitem {ray:capy}
Ray, D.: capy2vML: Provably-secure differentially-private machine learning training.
GitHub repository, GitHub, 2024.
\url{https://github.com/drcapybara/capy2vML}

\end{thebibliography}
\end{document}